\documentclass[twocolumn]{aa}
\usepackage{graphicx}
\usepackage{epsfig}
\begin{document}
\newcommand{\gsim}{\hbox{\rlap{$^>$}$_\sim$}}
\authorrunning{Dado, Dar \& De R\'ujula}
\titlerunning{On the Polarization of GRBs}
\title{On the Polarization of Gamma Ray Bursts\\
and their Optical Afterglows}
 \author{Shlomo Dado\inst{1} \and Arnon Dar\inst{1,2} \and
A. De R\'ujula\inst{2}}
\institute{Physics Department and Space Research Institute, Technion,
               Haifa 32000, Israel\\
           \and Theory Division, CERN, CH-1211 Geneva 23, Switzerland}
           \maketitle

\abstract{The polarization of the optical afterglow (AG)  of Gamma-Ray
Bursts (GRBs) has only been measured in a few instances at various times
after the GRB. In all cases except the best measured one (GRB 030329) the
observed polarization and its evolution are simple and easy to explain in
the most naive version of the ``Cannonball'' model of GRBs:  the
``intrinsic'' AG polarization is small  and the observations
reflect the ``foreground'' effects of the host galaxy and ours. The
polarization observed in GRB 030329 behaves chaotically, its understanding
requires reasonable but ad-hoc ingredients. The polarization of the
$\gamma$-rays of a GRB has only been measured in the case of GRB 021206.
The result is debated, but similar measurements would be crucial to the
determination of the GRB-generating mechanism.

\keywords{gamma rays: bursts---afterglow polarization: general}}


\section{Introduction}

Spectropolarimetric measurements of radiations from astronomical objects
are an important diagnostic tool of their production mechanism. Gamma Ray
Bursts (GRBs) are not an exception. For these phenomena one must
distinguish between two observables: the polarization of the ``prompt''
$\gamma$-rays of the GRB itself, which has been measured in just one  
instance (GRB 021206; Coburn and Boggs 2003) and the polarization of the
GRB afterglows (AGs), observed at optical frequencies in a handful of
cases (GRB 990510: Wijers et al.~1999; Covino et al.~1999; GRB 990712: Rol
et al.~2000; GRB 010222: Bjornsson et al.~2001; GRB 011211: Covino et
al.~2002; GRB 020405: Bersier et al.~2002; Masetti et al.~2003; Covino et
al.~2003a; GRB 020813: Barth et al.~2003; Covino et al.~2003b; Gorosabel  
et al.~2003; GRB 021004: Rol et al.~2003; Wang et al.~2003; GRB 030329:
Efimov et al.~2003;  Magalhaes et al.~2003; Covino et al.~2003c; Greiner  
et al. 2003).

In this paper we are primarily concerned with the polarization of optical
AGs, though we first comment on that of a GRB itself. We shall conclude
that, while (difficult) convincing measurements of the polarization of
GRBs would be decisive in establishing the {\it mechanism} generating
GRBs, the measurement of the polarization of AGs is unlikely (at least in
the CB model) to shed much light on their understanding.

\section{Observational results}

The first, and so far the only measurement of the prompt polarization
of a GRB was recently reported. Using the RHESSI (Reuven
Ramaty High Energy Solar Spectroscopic Imager) satellite, whose primary
mission is to look at the Sun in the $\gamma$-ray band, Coburn and Boggs
(2003) discovered the extremely bright GRB 021206, and measured a very
large linear polarization of its prompt $\gamma$-rays: $\Pi= (80\pm 20)\%$.
This polarization measurement has been criticized by Rutledge \&
Fox (2003), who obtain an upper limit $\Pi<4.1\%$ at 90$\%$
confidence from the same data. Boggs \& Coburn (2003) have criticized 
the critique, and announced a systematic reanalysis. We cannot judge
this controversy, nor its likely outcome.
 
The first polarimetric measurement of a GRB's optical AG was that of GRB
990123 (Hjorth et al.~1999) and was consistent with zero. 
The eight later optical-AG
observations (cited in the Introduction)  were positive detections of a
small linear polarization, typically $\Pi < 3\%$.

The recent measurements of the AG of GRB 030329 ---at 
a redshift $z=0.1685$ the nearest
GRB after GRB 980425--- made with an unprecedentedly frequent sampling in
time, show rapid variations in the magnitude and angle of its linear
polarization (Greiner et al.~2003). These variations do not appear to be
clearly
correlated with the observed deviations of the optical AG fluence from a
smooth behaviour.

With the exception of GRB 030329, for which the late-time AG data may be
``contaminated'' by its prominent associated supernova (Dado, Dar \& De
R\'ujula 2003c; Stanek et al.~2003, Hjorth et al.~2003), in all AGs wherein a
non-zero polarization was measured at late time, its value and position
angle were consistent, within errors, with the polarization induced by
dust in our own Galaxy along the line of sight to the GRB (e.g.~Covino et
al.~1999, 2003; Lazzati et al.~2003a).  In three cases for which there are
measurements at various times (GRBs 020405, 020813, 021004) the level of
polarization and/or the position angle evolved towards its late-time
constant value. These are strong indications that the intrinsic
polarization of the source tends with time to a small value.

\section{The polarization of the $\gamma$-rays of GRBs}

Two different mechanisms have been discussed as the possible dominant
sources of the $\gamma$-rays of GRBs: {\it Inverse Compton Scattering} 
(ICS) and {\it synchrotron radiation} (SR). The first mechanism naturally
results in a sizeable polarization, while the second one does not.

Shaviv \& Dar (1995) hypothesized that the $\gamma$-rays of a
GRB are produced by  ICS of
ambient photons  by electrons partaking in the bulk motion
of highly relativistic, very collimated jets. The jets
would be emitted in mergers of compact
stellar objects that lead to a gravitational collapse.  If the electrons'
Lorentz factor is $\gamma\sim 10^3$, target photons of energy
$\sim 1$ eV are upscattered to the observed energies,
higher by a factor $\sim \gamma^2$.
The outgoing photons are forward-collimated within a beam of
characteristic angular aperture $\sim 1/\gamma$. 
At an observer's angle $\theta$, the predicted polarization is:
\begin{equation}
\Pi(\theta,\gamma)\approx {2\;\theta^2\,\gamma^2/(1+\theta^4\,\gamma^4)},
\label{polSN}
\end{equation}
which, for the probable viewing angles,
$\theta\sim 1/\gamma$, is naturally large (Shaviv \& Dar 1995).

Contrariwise, the expected polarization 
vanishes (see e.g., Medvedev and Loeb 1999; 
Lyutikov, Parviev \& Blandford 2003) if 
the $\gamma$-ray generating mechanism is that of {\it fireball}
models: SR from shock-accelerated electrons moving in the
highly entangled magnetic field created by a relativistic
shell interacting with the circumburst medium (Katz 1994a,b). 
This is the case both for GRBs
produced by honest-to-goodness (i.e. spherical) fireballs
and for ``collimated fireballs''
viewed from the traditionally-adopted on--axis viewing position 
(Rhoads 1997, 1999; Sari, Piran \& Halpern 1999;
Frail et al.~2001; Berger et al.~2003a; Bloom et al.~2003). 

A GRB polarization large enough to be measurable, such as that
observed in GRB 021206 (if it is not in error) would
very clearly advocate in favour of ICS, as opposed to SR,
as the mechanism generating the $\gamma$-rays of a GRB
(Dar \& De R\'ujula, 2003; see also the later work\footnote{The
``sociological'' aspects of this work have been criticized
in De R\'ujula (2003).} of
Lazzati et al.~2004).
Interestingly, other authors reach the opposite conclusion.
Nakar, Piran \& Waxman (2003), for instance, state:
{ ``the recent detection of very high linear polarization...
suggests strongly that these $\gamma$-rays are produced by
synchrotron emission of relativistic particles.'' } 

SR from a power-law distribution of electrons
$dn_e/dE\sim E^{-p}$ in a {\it constant} magnetic field  
can produce a large polarization, $\Pi=(p+1)/(p+7/3)$,
that is $\approx 70\%$
for $p\approx 2.2$. But a collisionless shock acceleration of the 
electrons requires highly disordered and time varying magnetic fields 
(for a recent review see, e.g. 
Zhang \& Meszaros 2003, for a dissenting view on this
point, see Lyutikov, Pariev \& Blandford 2003).
Only under 
very contrived circumstances ---such as geometrical coincidences
and unnaturally ordered magnetic fields--- can shock 
models of GRBs  produce a large linear polarization. 
In our opinion, this is what recent 
articles (Eichler \& Levinson, 2003; Waxman, 2003; Nakar, Piran \&
Waxman, 2003) on the subject  show, although it is not
what they say.

Another problem with a SR origin of a
large polarization in shock models 
is that if synchrotron self-absorption is invoked 
to explain the low-energy spectral shape of GRBs, then 
the linear polarization is (Ginzburg and Syrovatski, 1969; Longair 1994)
$\Pi=3/(6p+3)< 12\%$,  parallel to the magnetic field.
Since most of the photons   
of GRB 021206 had energies below its peak energy 
(larger than 1 MeV), synchrotron 
self absorption and/or an entangled magnetic field should
result in  a  polarization $\Pi <12\%$. 
 
 \section{The polarization of AGs in fireball models}

Linear polarizations of the order of a few percent were proposed to arise
from causally-connected magnetic patches (e.g.~Gruzinov \& Waxman 1999), 
from homogeneous conical jets (Gruzinov 1999; Ghisellini \& Lazzati 1999; 
Sari 1999) and from structured jets viewed off-axis (Rossi, Lazzati \& Rees 2002).  
The observations have been
interpreted as evidence for a small intrinsic linear polarization of
the optical AG of GRBs (Lazzati et al.~2003a).

\section{The CB model}

In this model long-duration GRBs are produced in
core-collapse supernova (SN) events\footnote{Dar \& De R\'ujula (2003) 
argue that type Ia SNe are responsible for short-duration GRBs, 
while 
core-collapse SNe (Types Ib, Ic and II) are the progenitors
of long GRBs.}. An accretion disk around
the newly-collapsed core is supposed to be made
by stellar material that has not been efficiently
ejected. In analogy with processes seen to occur
in quasars and microquasars, the subsequent periods
of violent accretion of disk material lead to the
bipolar ejection of relativistic blobs of ordinary matter: 
{\it cannonballs.} Each CB generates one pulse of a GRB 
as it crosses and Compton up-scatters the ``ambient light'' 
surrounding the progenitor star. This model is very
successful in its very simple description of the properties of GRBs
(Dar \& De R\'ujula 2000a,b, 2003).

The CBs
initially expand (in their rest system) at a velocity comparable to,
or smaller than, the speed of sound in a relativistic plasma
($c/\sqrt{3}$), so that the jet opening angle ---subtended by
a CB's radius as observed from its emission point\footnote{We
are neglecting the initial CB's radius, presumably comparable
or not much bigger than that of the collapsed core of the parent star,
and thus entirely negligible by the time the GRB is emitted.})--- is 
$\alpha_j<\! 1/(\gamma_0\,\sqrt{3})$. 
An observer sees the ``Doppler-favoured'' jet, travelling
at a small angle $\theta={\cal{O}}(1/\gamma)$ relative to
the line of sight. 
Typically $\theta>\alpha_j$, so that the jet's
opening angle can be neglected and the observer's angle
is the {\it only} relevant one.
That is why the prediction of Eq.~(\ref{polSN}) for a narrow jet
is naturally incorporated in the CB model\footnote{To accommodate
the possibly observed large GRB polarization, 
Lazzati et al.~(2004) assume that the opening angle of
their ``fireball" ejecta is a few times $1/\gamma$, a completely
ad-hoc choice, in their case.}.

\section{AGs and their polarization in the CB model}

In the CB model the AG ---unlike the prompt
GRB--- is generated by electron SR
in the disordered magnetic mesh permeating a CB
(this was the only similarity between the CB model and
the fireball models, before the latter significantly evolved).

As a CB moves through the interstellar medium (ISM),
it gathers and scatters its constituent electrons and nuclei.
These generate within the CB chaotic magnetic fields
that accelerate all charged particles, in a ``Fermi"
acceleration process that was conjectured in Dar (1998)  
and has been numerically studied by Frederiksen et al.~(2003). 
Their results, based on ``first principles" (Maxwell's equations
and the Lorentz force) show that the process of acceleration
does not involve the formation of any shocks, contrary to the
customary basic assumption of fireball models.

The AG is the synchrotron radiation from the accelerated electrons,
in the CBs magnetic field, whose magnitude is predictable
(Dado et al.~2002a). This model is very successful in its
description of the properties of X-ray, optical
(Dado et al.~2002a) and radio AGs (Dado et al.~2003a).

\subsection{Intrinsic polarization}

The naive expectation is that, since the magnetic field within a CB is 
disordered, the {\it intrinsic} AG polarization ought to be small,
since it would result from a fractional ``order'' in a disordered field.
The currents induced by the ISM ---which in a CB's rest system 
impinges onto the CB as a one-directional relativistic wind---
may depend on the ISM's varying density, and have a 
non-vanishing ``convective" component. The CBs are viewed
at an angle relative to their direction of motion, so that symmetry
considerations do not prevent the possible existence of
a small and time-dependent intrinsic polarization.
 
It would be very difficult, and arguably
uninteresting, to estimate the precise magnitude of a CB's 
intrinsic polarization. Here we deal with this problem in the most expedient
fashion: setting the intrinsic polarization to zero and studying 
phenomenologically whether this simplest ansatz is tenable.
The case of GRB 030329, with its very time-dependent polarization, 
will force us to envisage the possibility (but not the unavoidable
conclusion) of the existence of small but non-vanishing intrinsic 
polarizations.

\subsection{Extrinsic (or foreground) polarization}

The subject of the AG polarization is rendered messier by unavoidable,
{\it extrinsic} time-varying contributions, expected and observed
to be of the same order of magnitude as the measured polarization
levels. These effects are induced by dust along the line of sight to the GRB.
The total extrinsic polarization 
results from the cumulative effects of the dust in the GRB's host
galaxy and in our Galaxy. In the CB model, moreover, the
host-induced contribution to the
polarization is time-dependent, since the CBs responsible
for the GRB and for its AG travel distances of the order
of kiloparsecs during the time the AG is observed. In this journey,
CBs ought to depart from the dustier central star-forming region of 
the host galaxy, where the event originates. They may also exit a
 ``super-bubble", to encounter enhanced and varying dust concentrations. 
 
 \subsection{Total polarization}

The polarization of the AG light from each of its uncorrelated sources
(intrinsic to the CB, the underlying SN and that
induced by the magnetized
ISM dust of the host galaxy and ours) is linear and small. 
Let $Q_i$ and $U_i$ be the customary (normalized) Stokes' parameters,
characterizing linearly, partially polarized light,
with $i$ an index running over the three sources. Let $Q=\Sigma\,Q_i$
and $U=\Sigma\,U_i$. The cumulative degree of linear polarization and 
its angle are simply $\Pi\simeq (U^2+Q^2)^{1/2}$ and $\tan 2\chi\simeq U/Q$.

Only the Galactic contribution to the AG polarization is fixed in
magnitude, angle and time. The polarization of the SN light is time
dependent and may be approximated by that of SN1998bw (Patat et al. 2001).
A priori we cannot tell whether a potential intrinsic polarization is a
function of time. The time dependence of the host-galaxy's contribution
---due to the CB's motion in the host galaxy--- requires a more detailed
discussion.

\subsection{The polarization induced by the host galaxy}

Let $\gamma_0$ be the original Lorentz factor of a GRB's
CBs\footnote{Unless otherwise stated, we approximate the
theoretical form of the AG by the contribution of a single
dominant CB.}  and $\gamma(t)$ its value after an observer's time
$t$, diminishing as the CBs decelerate as a consequence
of their interaction with the ISM ($t=0$ is the GRB's
trigger time). We have repeatedly reported in the
literature the explicit form of the function $\gamma(t)$,
a function of $\gamma(0)$
and $x_\infty$, a characteristic distance for the 
CB's slowdown (see, e.g.~Dado et al.~2002). In the approximation
of a constant-density ISM:
\begin{eqnarray}
\gamma&=&\gamma(\gamma_0,\theta,x_\infty;t)
= {B^{-1}} \,\left[\theta^2+C\,\theta^4+{1/C}\right]\, ,\nonumber\\
 C&\equiv&
\left[{2/
\left(B^2+2\,\theta^6+B\,\sqrt{B^2+4\,\theta^6}\right)}\right]^{1/3}\, ,
\nonumber\\
 B&\equiv&
{1/ \gamma_0^3}+{3\,\theta^2/\gamma_0}+
{6\,c\, t/ [(1+z)\, x_\infty]}\, ,
\label{cubic}
\end{eqnarray}
with $z$ the redshift of the host galaxy.

Let $\delta(t)\approx 2\,\gamma(t)/[1+(\gamma(t)\,\theta)^2]$
be the Doppler factor by which the energy of a photon
is boosted by the CB's motion, at the viewing angle $\theta$
towards the observer (the approximation is for large
$\gamma$, small $\theta$, the domain of interest).
An observer's time interval and the corresponding CB's travelled
distance are related by $dx/c=dt\,\gamma(t)\,\delta(t)/(1+z)$. The very large 
typical values of the coefficient multiplying $dt$,
of ${\cal{O}}(10^6)$ for small $t$, imply that CBs travel
for kiloparsec distances in months of observer's time.
The integrated distance travelled by CBs since their emission
is $x=x_{\infty}\,[1/\gamma(t)-1/\gamma(0)]$, typically of
order a kiloparsec at $t$ of order one week.

The contribution of the host galaxy to the polarization
may be a complicated function of time, since
the CBs are travelling for long distances and the line of sight
from the CB to the observer is changing in length and in
angular position in the sky. Both the degree and the direction of 
the induced polarization may be quite variable, since they depend 
on column-density-like integrals along the line of sight.
Here we explore the simplest possibility by assuming 
for the host galaxy's contribution a
constant polarization direction, $\chi_{_{H}}$, 
and arguing for an approximately
exponentially-varying degree of polarization, $\Pi_{_{H}}$.

Since GRB progenitors are observed to populate the dense,
central, actively star-forming regions of their host galaxies
(Djorgovski et al.~2003), we shall make the approximation that the 
density
of the ISM {\it dust} away from the parent SN decreases 
exponentially\footnote{There is no contradiction with the
constant density used in deriving Eq.~(\ref{cubic}), which
refers to the bulk of the ISM at kpc distances and not the 
dust contamination at shorter distances.}
with distance, with a characteristic fall-off length
$x_0$ (we are avoiding the term ``hight'' because
the CBs would typically travel in a slant direction relative
to the normal to a galaxy's disk). The integrated 
host-galaxy column-density in the observer's direction
(and the subsequent polarization) are then of the form:
\begin{equation}
\Pi_{_{H}}(t)=\Pi_0\,Exp\left[-{x(t)\over x_0}\right]\equiv \Pi_0\,
Exp \left[{b \over \gamma_0}-{b\over \gamma(t)} \right] \, ,
\label{pola}
\end{equation}
where $b\propto x_\infty/x_0$ is a parameter to fit.
Because we are assuming a fixed polarization angle, $\chi_{_{H}}$, the Stokes 
parameters of the host-induced effect vary in the same way
as $\Pi_{_{H}}$ does:
\begin{eqnarray}
Q_{_{H}}(t)&=&Q_0\,\left[{b \over \gamma_0}-{b\over \gamma(t)} \right] \, ,
\nonumber\\
 U_{_{H}}(t)&=&Q_{_{H}}(t)\,\tan 2\chi_{_{H}}
\label{host}
\end{eqnarray}

\subsection{The fitting procedure}

The data on the time evolution of the optical and radio AG fluence at
various frequencies is typically much more abundant than the data on the
AG polarization. Given this, we first fit the fluence data, thereby
extracting the parameters ($\gamma_0$, $x_\infty$ and $\theta$) that
determine the function $\gamma(t)$ for each individual GRB. The way these
fitting is performed is described in minute detail in Dado et al.~(2003a).
We subsequently fit the observed Stokes parameters $Q$ and $U$ to the sum
the host-induced functions of Eq.~(\ref{host}) and their constant
Galactic-induced values (except for GRB 030329, these values are those of
the late-time measurements, introduced in the fits with their
corresponding uncertainties). The polarization-fit parameters are $Q_0$,
$b$, and $\chi_{_{H}}$.  In the case of GRB 030329 and GRB 021004, the
contribution of the two CBs are weighted according to their relative
contribution to the optical light curves as function of time.

\section{GRBs 020405, 020813, 021004 and 030329}

These GRBs are the 
ones for which there is data on the time-dependence of the polarization
of the AG. Their parameters describing our best fits to the AG fluence 
and polarization are given in Table I, where we have reported the
polarizations levels and angles, rather than the Stokes parameters.

\vskip .5cm
\noindent
{\bf Table I:} Inputs and parameters of the CB-model description of 
the AG fluence and polarizations of GRBs 020405, 020813, 021004 and 030329.
$\gamma_0$, $x_\infty$ and
$\theta$ describe the AG fluence and determine  
$\gamma(t)$ via Eq.~(\ref{cubic}). The host-galaxy effect is described by
the initial polarization
level $\Pi_0$, its exponential decay constant $b$, and the 
polarization angle $\chi_{_{H}}$. But for the last GRB, 
whose AG is fit with two CB contributions, the Galactic (or late-time) 
parameters,  $\Pi_{_{G}}$ and $\chi_{_{G}}$,  are inputs.
\begin{table}[h]
\normalsize
\begin{tabular}{|l|c|c|c|c|c|}
\hline
Parameter & 0405 & 0813 & 1004 & 0329 \\
\hline
$z$                & 0.69    &  1.2545    & 2.328       &    0.1685    \\
$\gamma_0$         & 645     &  1173      & 1403;~1259  &  1037;~1606  \\
$ x_{\infty}$[Mpc] & 0.31    &  0.54      & 0.025;~0.62 &  0.033;~0.37 \\
$\theta$ [mrad]    & 0.42    &  0.14      & 1.47;~1.47  &  2.20;~2.30  \\
$\Pi_0\,[\%]$            & 1.93    &  4.45      & 1.036       &  4.29        \\
$ b $              & 19.02   &  2005      & 128         &  4342        \\
$\chi_{_{H}}$ [deg]& 144     &  145       & 138         &   121        \\
$\Pi_{_{G}}\,[\%]$   & 1.10    &  0.55      & 0.64        &   0.52       \\
$\chi_{_{G}}$ [deg] & 24.2    &  177       & 11.4        &  51.6        \\
\hline

\end{tabular}
\end{table}

\begin{figure}
\hskip 2truecm
\vspace*{0.3cm}
\hspace*{-1.6cm}
\epsfig{file=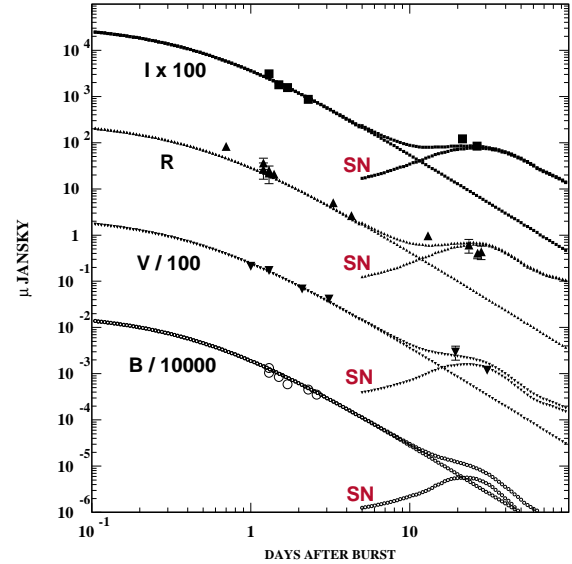, width=8.5cm}
\caption{CB model  fit to the
measured I, R, V, and B-band AG of GRB 020405.
The various bands are scaled for presentation
(see Dado et al.~2002a for details).
The observations are not corrected
to eliminate the effect of extinction, thus
the theoretical contribution  from a SN1998bw-like supernova
was dimmed by the known extinction in the Galaxy and our consistently
estimated extinction in the host. The contribution of the host galaxy,
subtracted from the data by the HST observers, is not included in the fit.
\label{WB405}}
\end{figure}

\begin{figure}
\hskip 2truecm
\vspace*{0.3cm}
\hspace*{-1.6cm}
\epsfig{file=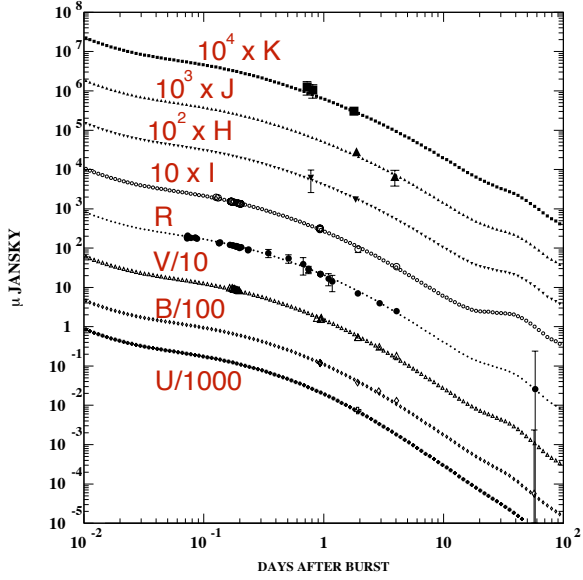, width=8.5cm}
\caption{
Comparison between the observations in the
K, J, H, I, R, V, B and
U bands of the optical afterglow of GRB 020813 
(Covino et al.~2003b, Li et al.~2003, Urata et 
al.~2003, and Gorosabel et al.~2003),
and the CB model
fit assuming one dominant CB (for details see e.g.
Dado et al.~2003b). 
 The various bands are rescaled for presentation.
\label{WB813}}
\end{figure}

\begin{figure}[]
\vskip .5cm
\vspace*{- .4cm}
\epsfig{file=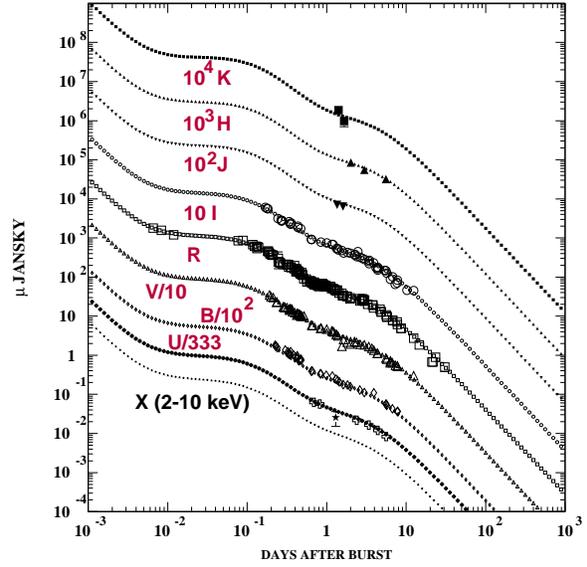, width=8.5cm}
\caption{The NIR--optical observations of the AG of GRB 021004
and the fit for two CBs with different parameters, corrected for extinction.
The ISM density is a constant plus a
``wind'' contribution decreasing as $\rm 1/r^2$.  The various bands
are scaled for presentation.  The data are those reported to date,
in GCN notices (recalibrated with the observations of Henden et
al. 2002), and in Bersier et al.~(2002); Holland et al.~(2002).
The  host-galaxy's contribution
was subtracted from the late-time I, R and V data, where it is
significant.  The X-ray Datum is from Sako et al.~(2002).
\label{fone}}
\end{figure}

\begin{figure}
\hskip 2truecm
\vspace*{0.3cm}
\hspace*{-1.6cm}
\epsfig{file=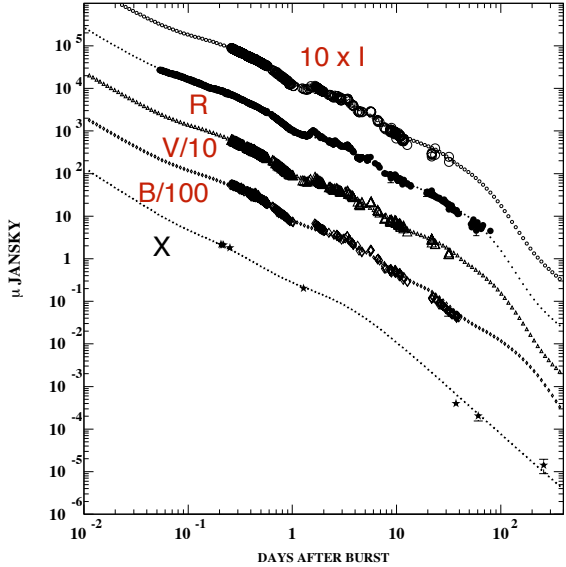, width=8.5cm}
\caption{The NIR--optical and X-ray observations of the AG of GRB 030329
and a broad-band fit for two CBs with different parameters,  described 
in the Dado et al.~(2002). The ISM density is assumed to be a 
constant plus a ``wind'' contribution decreasing as $ 1/r^2$.  The 
various bands are scaled for presentation. The fit is to
the X-ray data of RXTE (Marshall \& Swank, 2003;  Marshall, Markwardt \& 
Swank, 2003) and XMM-Newton (Tiengo et al.~2003) and many other NIR-optical 
measurements, recalibrated by Lipkin et al.~(2003 and references therein);
as well as the radio data of Sheth et al.~(2003) and Berger et al.~(2003b).
 The  host-galaxy's contribution
was neglected.  The individual bands have been rescaled for
clarity.
\label{WB329}}
\end{figure}

The CB-model fits to the NIR-optical  AG light curves (which in all cases 
but that of GRB 020405 are a subset of a broader-band fit
including radio data) are given in Figs.~(\ref{WB405}) to (\ref{WB329}).
The fit to GRB 020813 is new, all others have been previously
published or posted in the Archives (020405, 021004, 030329:
Dado et al.~2002b, 2003b, 2004, respectively).
The fits to the observed AG polarization and angle are shown in 
Figs.~(\ref{POL0208133}) to (\ref{329b}).

\section{Discussion and conclusions}

\begin{figure}[t]
\vspace{-.5cm}
\hskip 2truecm
\hspace*{-2.1cm}
\epsfig{file=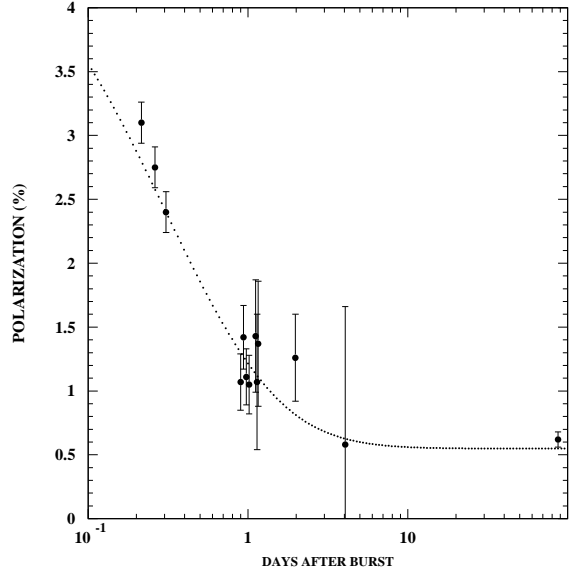, width=8.5cm} \\
\vspace{-0.5cm}
\caption{Comparison between 
the linear polarization of the optical AG of GRB 020813 
measured by  Gorosabel et al.~2003 and the CB model fit assuming the 
polarization is {\it extrinsic}: produced by 
scattering of light by dust in the ISM along the line of sight in the host 
galaxy and in the Milky Way. The point at 100 days is the polarization
of starlight in the Milky Way along the line of sight.}
\label{POL0208133}
\end{figure}

\begin{figure}[t]
\vspace{-.5cm}
\hskip 2truecm
\hspace*{-2.1cm}
\epsfig{file=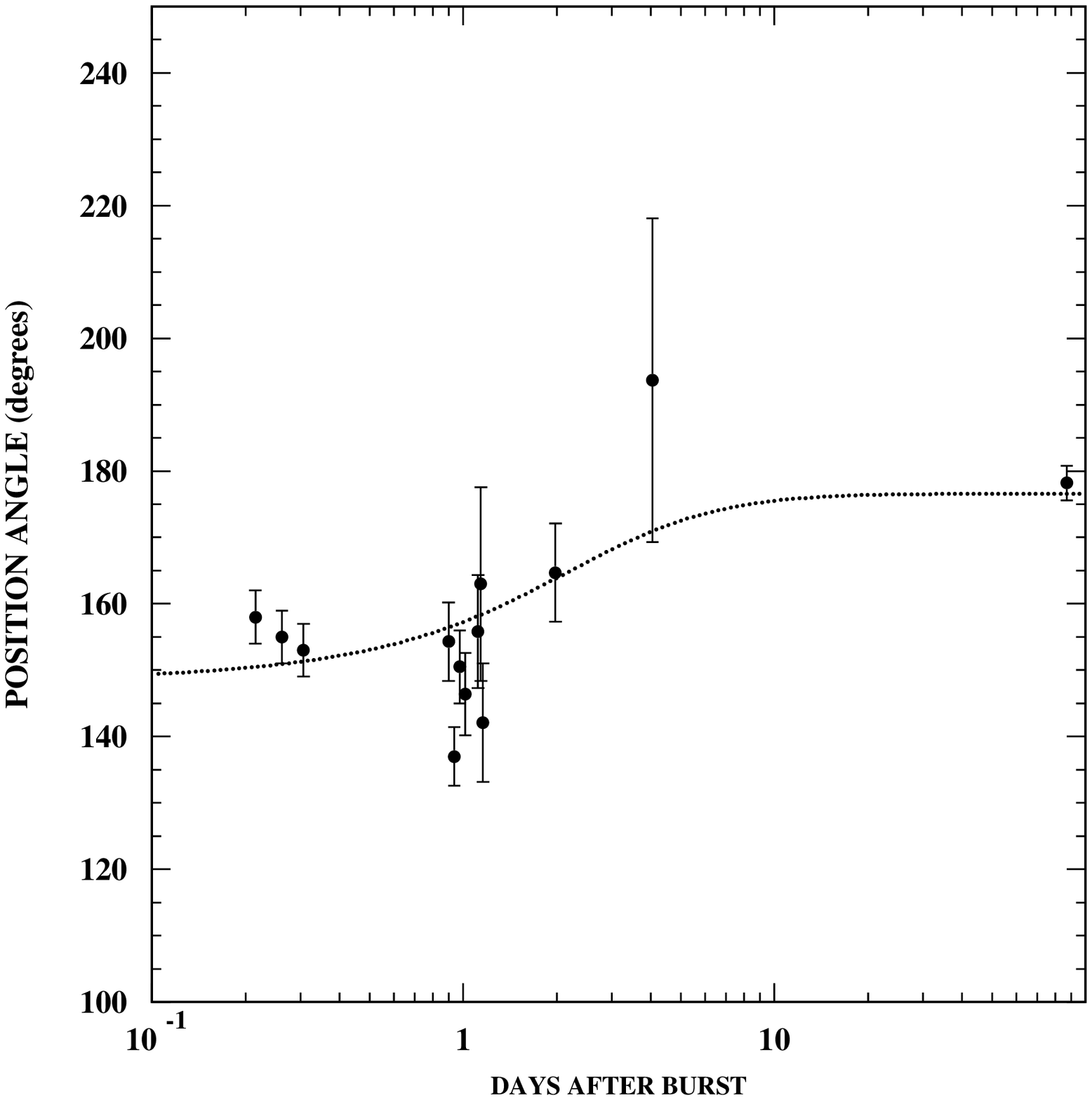, width=8.5cm} 
\vspace{-0.5cm}
\caption{Comparison between the position angle of 
the linear polarization of the optical AG of GRB 020813 
measured by Gorosabel et al.~(2003)
and the CB model fit assuming the linear polarization is  
{\it extrinsic}. The point at 100 days is the position angle
of the polarization of starlight in the Milky Way along the line of sight.}
\label{POS0208133}
\end{figure}

\begin{figure}[t]
\vspace{-.5cm}
\hskip 2truecm
\hspace*{-2.1cm}
\epsfig{file=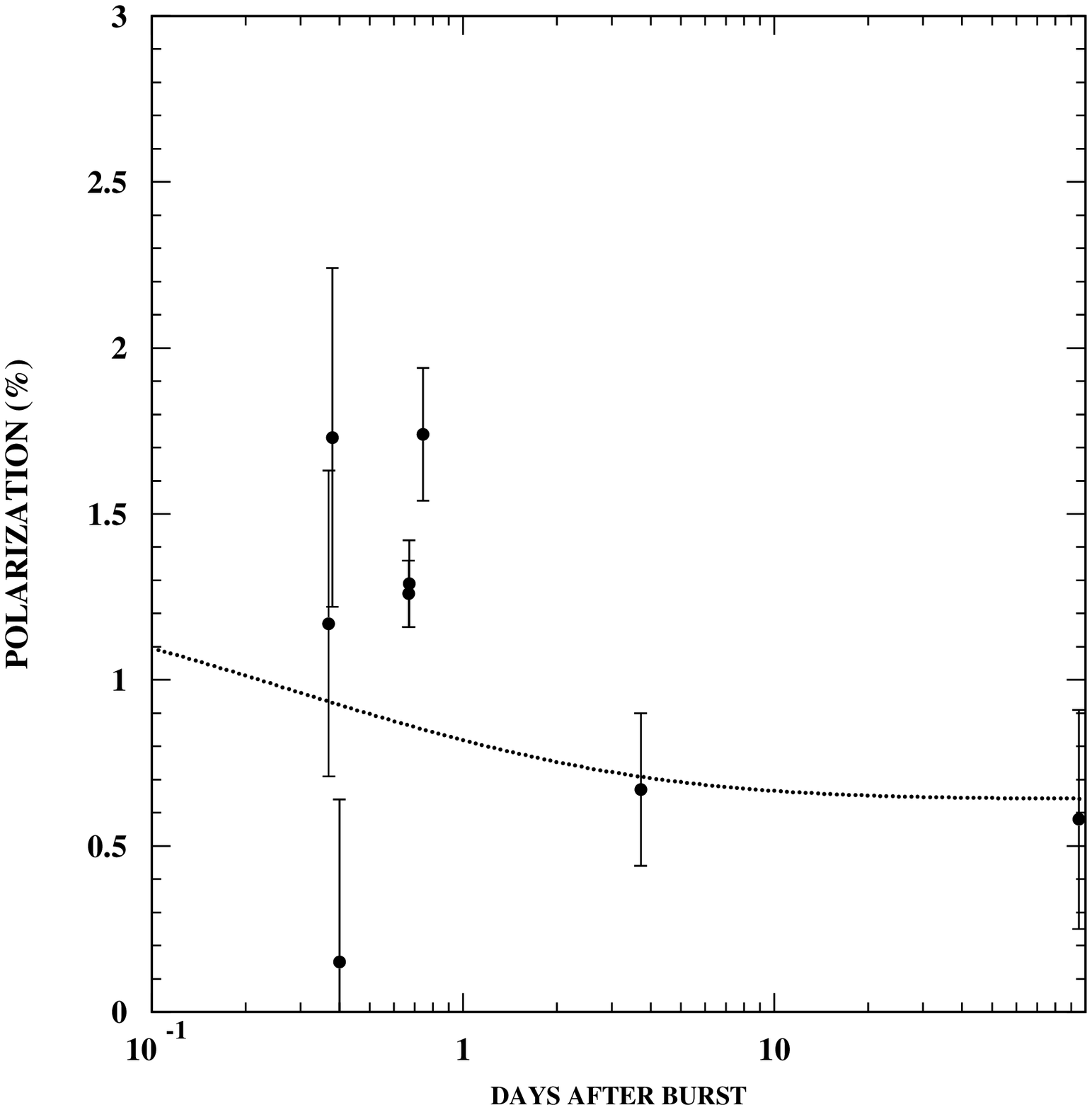, width=8.5cm} \\
\vspace{-0.5cm}
\caption{Comparison between 
the linear polarization of the optical AG of GRB 021004 
measured by Rol et al.~(2003) and Wang et al.~(2003),
and the CB model fit assuming the polarization is {\it extrinsic}.
The point at 100 days is the polarization
of starlight in the Milky Way along the line of sight.}
\label{004a}
\end{figure}

\begin{figure}[t]
\vspace{-.5cm}
\hskip 2truecm
\hspace*{-2.1cm}
\epsfig{file=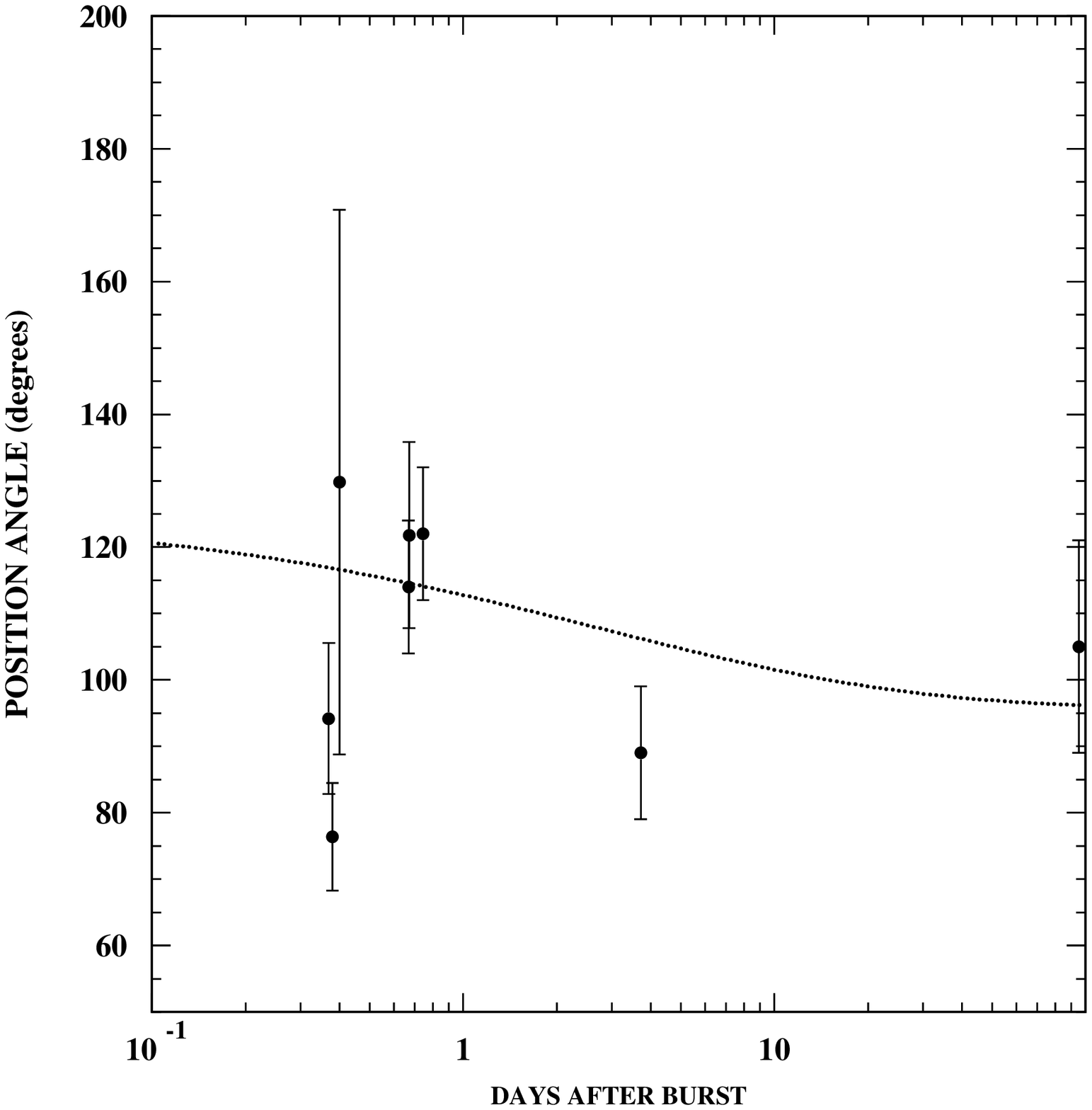, width=8.5cm} \\
\vspace{-0.5cm}
\caption{Comparison between the position angle of 
the linear polarization of the optical AG of GRB 021004
measured by Rol et al.~(2003) and Wang et al.~(2003),
and the CB model fit assuming the linear polarization is  
{\it extrinsic}. The point at 100 days is the position angle
of the polarization of starlight in the Milky Way along the line of sight.}
\label{004}
\end{figure}

\begin{figure}[t]
\vspace{-.5cm}
\hskip 2truecm
\hspace*{-2.1cm}
\epsfig{file=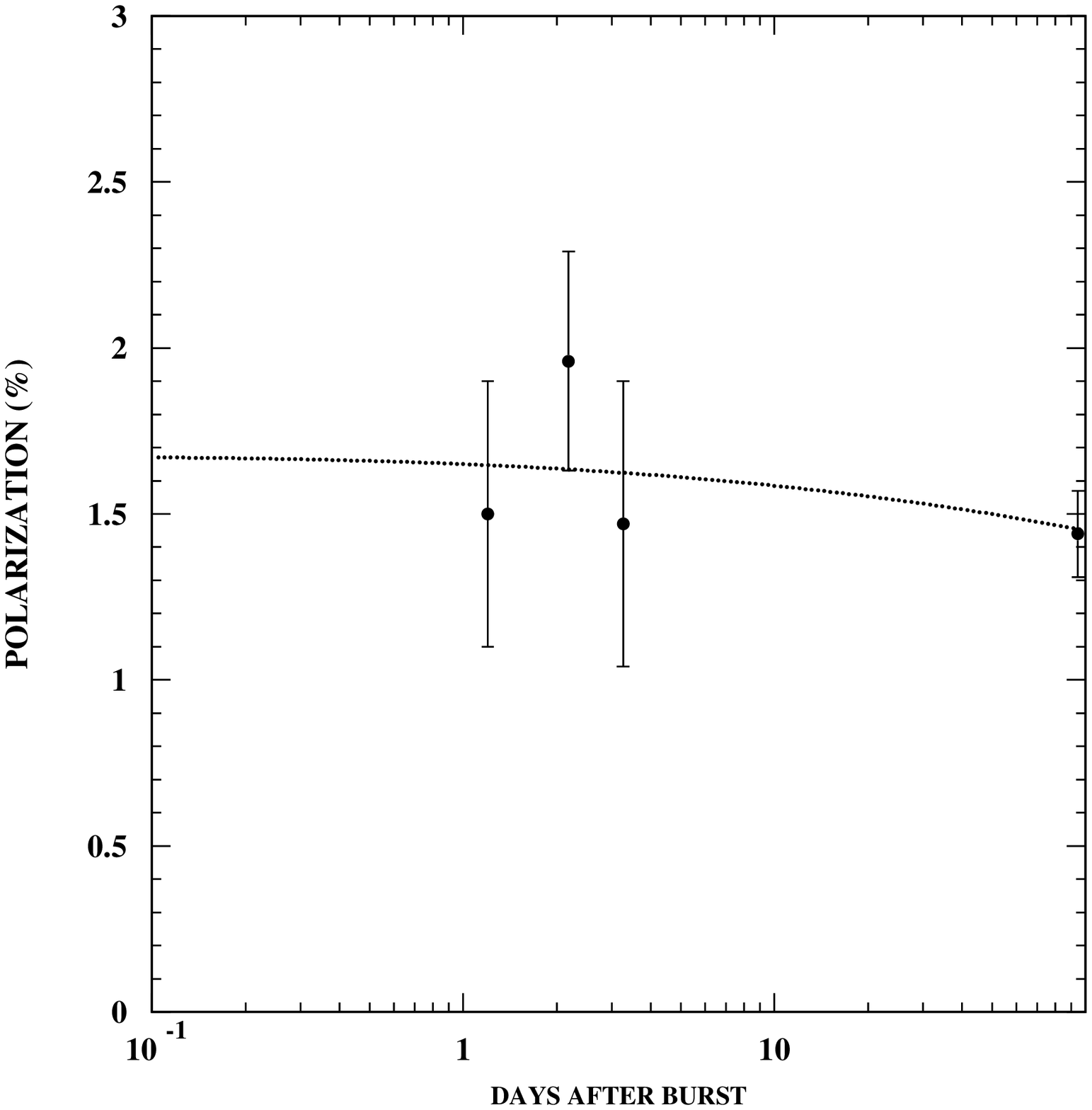, width=8.5cm} \\
\vspace{-0.5cm}
\caption{Comparison between 
the linear polarization of the optical AG of GRB 020405
measured by Bersier et al.~(2002); Masetti et al.~(2003); Covino et
al.~(2003a),
and the CB model fit assuming the polarization is {\it extrinsic}. 
The point at 100 days is the polarization
of starlight in the Milky Way along the line of sight.}
\label{405a}
\end{figure}

\begin{figure}[t]
\vspace{-.5cm}
\hskip 2truecm
\hspace*{-2.1cm}
\epsfig{file=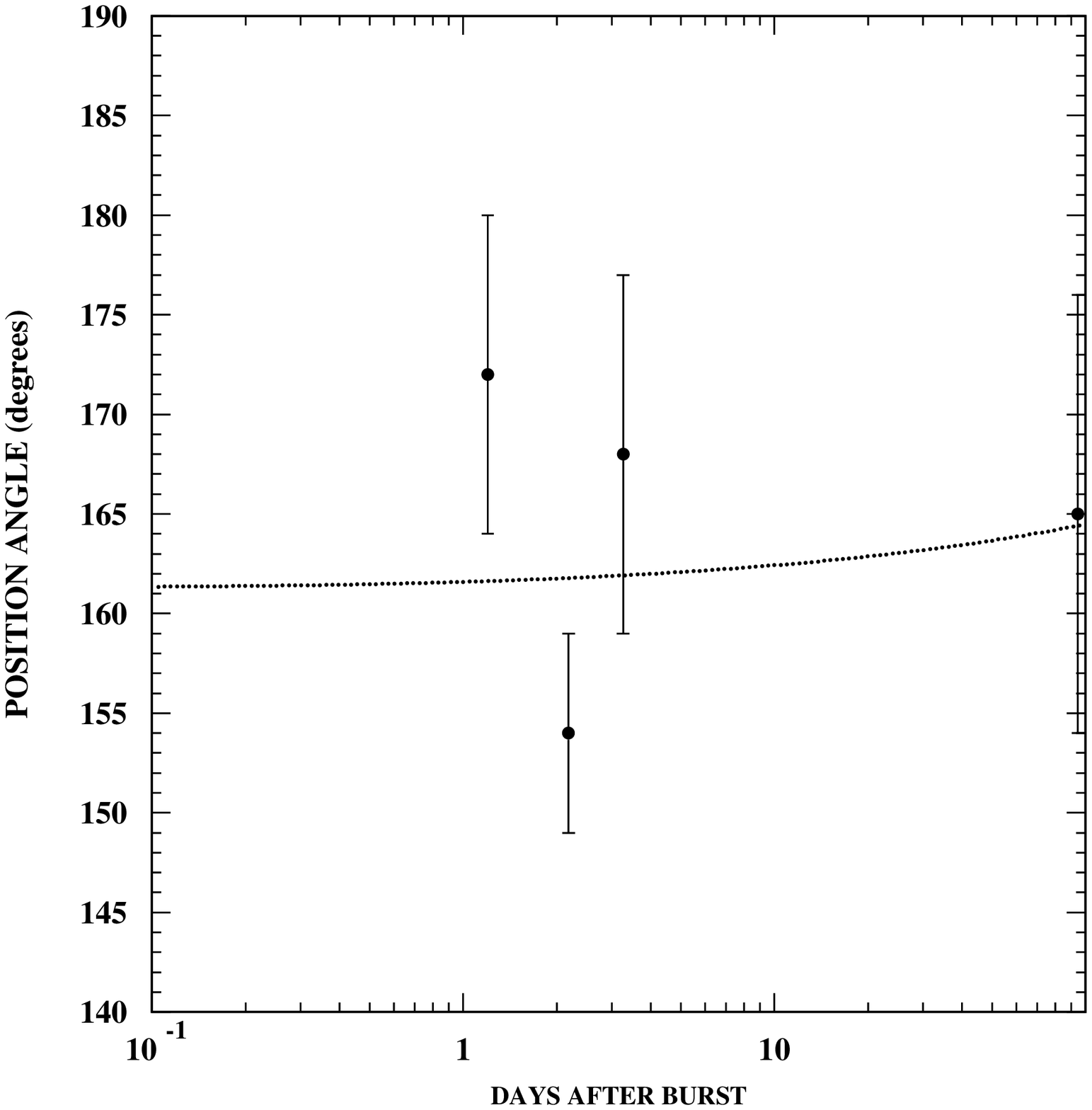, width=8.5cm} \\
\vspace{-0.5cm}
\caption{Comparison between the position angle of 
the linear polarization of the optical AG of GRB 020405
measured by  Bersier et al.~(2002); Masetti et al.~(2003); Covino et
al.~(2003a),
and the CB model fit assuming the linear polarization is  
{\it extrinsic}. The point at 100 days is the position angle
of the polarization of starlight in the Milky Way along the line of sight.}
\label{405b}
\end{figure}

\begin{figure}[t]
\vspace{-.5cm}
\hskip 2truecm
\hspace*{-2.1cm}
\epsfig{file=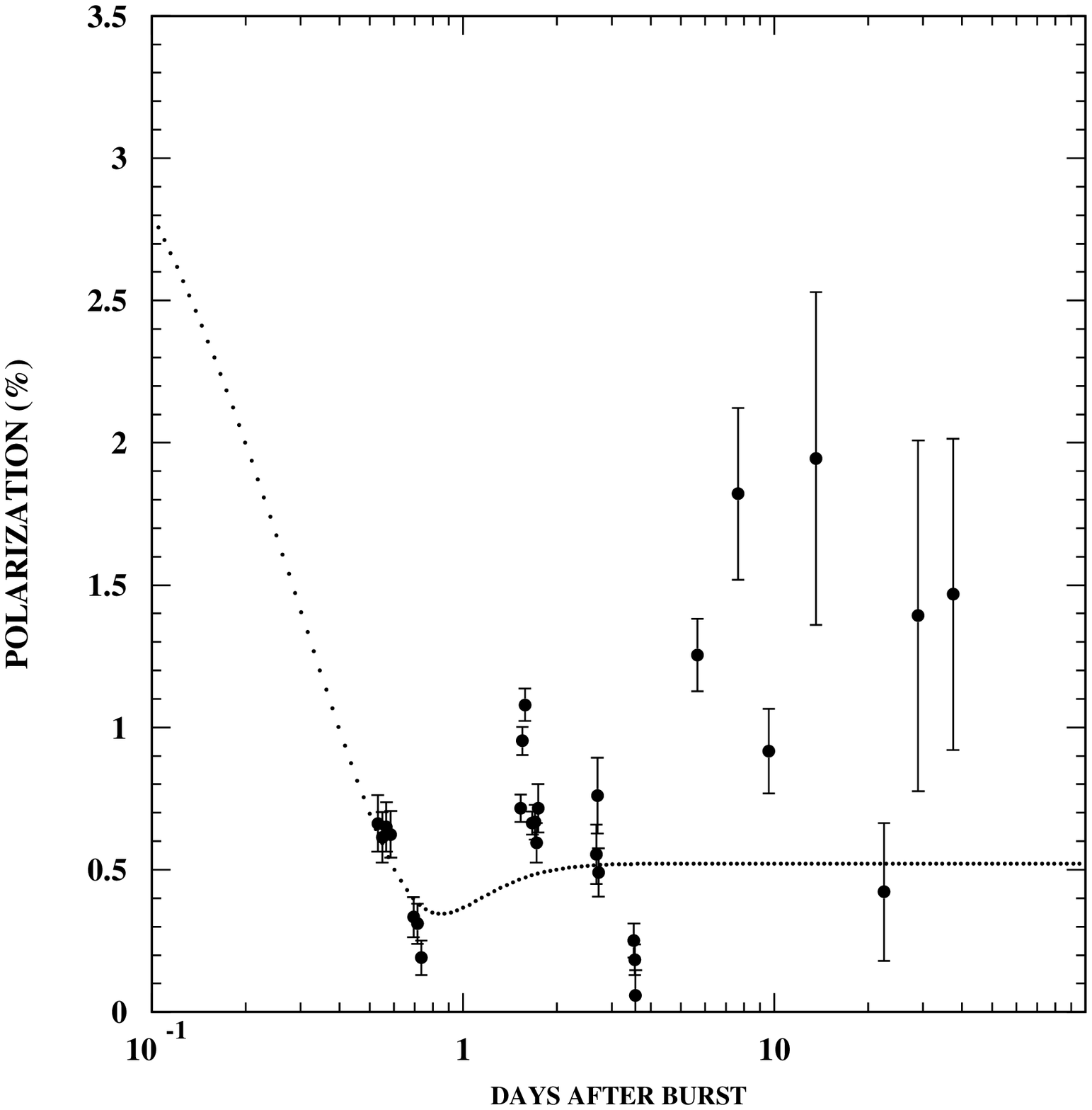, width=8.5cm} \\
\vspace{-0.5cm}
\caption{Comparison between 
the linear polarization of the optical AG of GRB 030329 
measured by Efimov et al.~(2003), Magalhaes et al.~(2003),  Covino et 
al.~(2003c) and  Greiner et al.~(2003),          
and the CB model fit assuming no {\it intrinsec} polarization
and a host-induced polarization simply described by Eq.~(\ref{pola}).
The ansatz clearly fails.}
\label{329a}
\end{figure}

\begin{figure}[t]
\vspace{-.5cm}
\hskip 2truecm
\hspace*{-2.1cm}
\epsfig{file=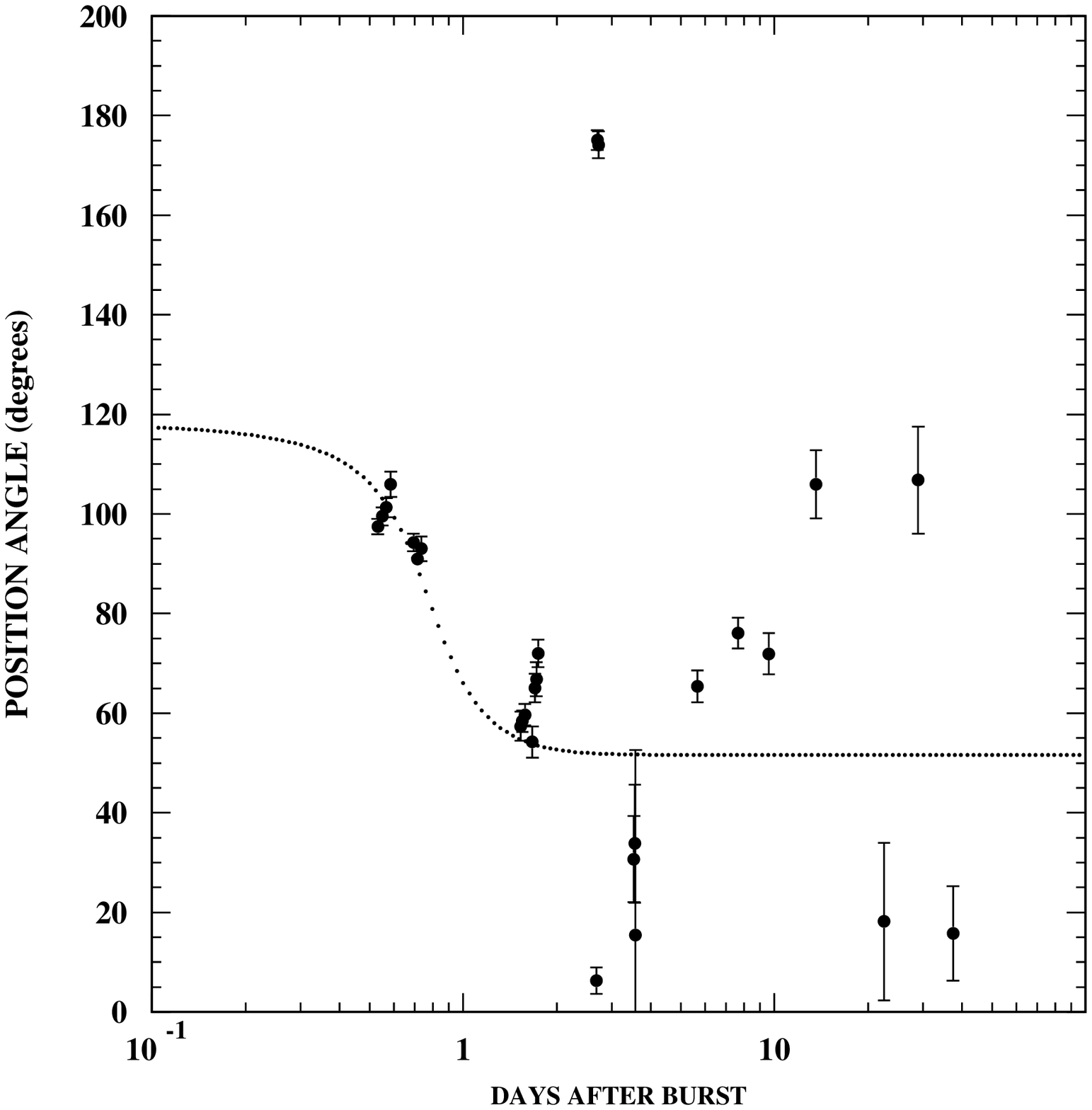, width=8.5cm} \\
\vspace{-0.5cm}
\caption{Comparison between the position angle of 
the linear polarization of the optical AG of GRB 030329
measured by Efimov et al.~(2003), Magalhaes et al.~(2003),  Covino et 
al.~(2003c) and  Greiner et al.~(2003),
and the CB model fit assuming no {\it intrinsic} polarization
and a host-induced polarization simply described by Eq.~(\ref{pola}).
The ansatz clearly fails.}
\label{329b}
\end{figure}

Examining the results shown in Figs.~(\ref{POL0208133}) to (\ref{329b}), 
we conclude that,
for GRBs 020405, 020813 and 021004, the fits ---which 
did not include an {\it intrinsic} polarization of the light emanating from the CBs---
are good enough to conclude that this contribution vanishes within errors.
As required in that case, the polarization and angle tend at late times to
the Galactic foreground values. This is particularly convincing for GRB
020813. The most naive expectation ---that the observed polarization simply
reflects the foreground effects of the host galaxy and ours--- is vindicated.

The case of GRB 030329, for which the data are particularly precise
and abundant, brings havoc to the previous clear and simple conclusion.   
This is the case even if one neglects the data after day $\sim 6$,
probably contaminated by the parent SN, which is particularly prominent in
this instance. 

In the CB-model, the explanation for the ``humps" in the AGs of some GRBs, 
such as GRBs 970508 and 000301c (Dado et al.~2002a)
and 030329 itself (Dado et al.~2004), is simple:
the optical fluence $F_\nu(t)$ is a direct and {\it quasi-local} tracer of
the density of the ISM through which a CB travels: spatial changes in density
translate into temporal changes in fluence. This statement
is not  as sterile as it sounds, for it permits the 
extraction of the circumburst density and its radial profile 
from the time-dependence of the early AGs and, very satisfactorily,
the magnitude and $\sim 1/r^2$ profile of the result are those expected 
from observations
of ``winds'' of ``pre-supernova'' massive stars (Dado et al.~2003b).

\begin{figure}
\epsfig{file=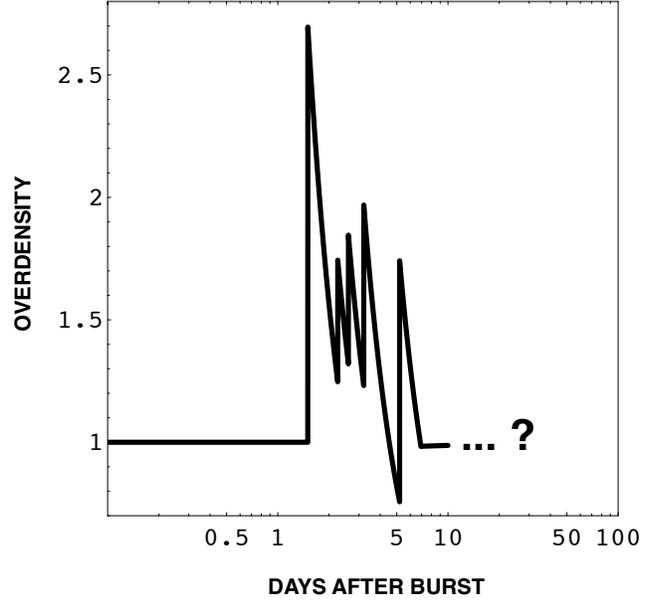, width=8.5cm}
\caption{The overdensity (relative to a smoothly varying function)
of the ISM traversed by the CBs of GRB 030329 (Dado et al.~2004),
shown as a function of observer's time, for comparison with the
polarization results of Figs.~(\ref{329a},\ref{329b}).
The fit to the AG does not determine the density for $t>10$
days, a time at which the observations are dominated by the
associated SN.}
\label{overdensity}
\end{figure}

In the case of GRB 030329, the observed deviations of the AG light
curves from smoothly-varying functions are attributed 
to density fluctuations encountered by the CBs at the
time they exit from the superbubble in which the explosion took
place. The size and shape of these density fluctuations can be
explicitly extracted from the data: they are a series of
density jumps followed by $\propto 1/r^2$ declines, as befits
the remnants of the explosions that created the superbubble
(Dado et al.~2004). Their magnitude and shape (as functions
of time) are shown in Fig.~(\ref{overdensity}).

Fluctuations of the host's ISM density though which a CB travels may also 
cause the polarization fluctuations observed in GRB 030329. This may be
a foreground ``integral'' effect (induced by the varying amounts and field 
directions of magnetized dust in the complicated density profile
along the line of sight to the observer). It  may 
also be a local ``intrinsic" effect: the magnetic field within a CB may not be, at the
few percent level, totally chaotic. It may be influenced by fluctuations
in the density of the ISM particles that impinge into it and generate its
magnetic structure. There appears to be no clearly convincing correlation
between the fluctuations in the fluence and those in the polarization,
though the periods of rapid variability coincide: compare 
Figs.~(\ref{329a}) and (\ref{329b}) to Fig.~(\ref{overdensity}).
This inclines the balance somewhat
in favour of an integral foreground effect, as if indeed there were
no intrinsic polarization in the radiation emanating from the CBs,
as expected for synchrotron radiation in a thoroughly disordered magnetic
field.

Admittedly, the considerations of the previous paragraph are not simple
and robust. They drive the conclusion that, in the CB model, no much is to
be learned from the polarization of optical AGs. This is in contrast to
the polarization of the $\gamma$-rays of a GRB, which, we contend, is
crucial for deciding what the GRB-generating mechanism is: inverse Compton
scattering if the polarization is measurably large.
 
\vskip .3cm

{\bf Acknowledgments}

The authors are grateful to J. Greiner for providing tabulated
data on the polarization of the AG of GRB030329. 
S. Dado and A. Dar are indebted to the theory Division of CERN for
hospitality. A. De R\'ujula is indebted to the Physics Department and
Space Research Institute of the Technion for its hospitality.  This
research was supported in part by the Helen Asher Space Research Fund for
research at the Technion.

\newpage
\pagebreak

\end{document}